\documentclass[pra,onecolumn,floatfix,superscriptaddress,longbibliography,notitlepage, nofootinbib]{revtex4-1}
\pdfoutput=1

\usepackage[utf8]{inputenc}
\usepackage{braket}
\usepackage{amsmath}
\DeclareMathOperator{\Tr}{Tr}
\usepackage{dcolumn}   
\usepackage{bm}        
\usepackage{amssymb}
\usepackage{mathtools}
\usepackage{tikz}
\usepackage{qcircuit}
\usepackage{appendix}
\usepackage{hyperref}
\usepackage{subcaption}
\usepackage{amsmath,amsfonts,amssymb}
\usepackage{mathtools}
\usepackage{thmtools,thm-restate}
\usepackage{algorithm}
\usepackage{mathrsfs}
\usepackage{tikz}
\usepackage{subcaption}
\usepackage{pgfplots}
\usetikzlibrary{3d}
\usepackage{tkz-euclide}

\usepackage{algpseudocode}

\usepackage{ulem}
\newcommand{\tcw}[1]{\textcolor{red}{#1}}
\newcommand{\tw}[1]{\textcolor{red}{#1}}

\newcommand{\Ibb}{\mathbb{I}}

\newtheorem{definition}{Definition}

\newtheorem{lemma}{Lemma}

\usetikzlibrary{arrows.meta}

\def\BibTeX{{\rm B\kern-.05em{\sc i\kern-.025em b}\kern-.08em
    T\kern-.1667em\lower.7ex\hbox{E}\kern-.125emX}}

\begin{document}
\title{Estimation of Nonlinear Physical Quantities By Measuring Ancillas}
\author{Nhat A. Nghiem}
\email{nhatanh.nghiemvu@stonybrook.edu}
\affiliation{Department of Physics and Astronomy, State University of New York at Stony Brook, Stony Brook, NY 11794-3800, USA}
\affiliation{C. N. Yang Institute for Theoretical Physics, State University of New York at Stony Brook, Stony Brook, NY 11794-3840, USA}

\author{Tzu-Chieh Wei}
\affiliation{Department of Physics and Astronomy, State University of New York at Stony Brook, Stony Brook, NY 11794-3800, USA}
\affiliation{C. N. Yang Institute for Theoretical Physics, State University of New York at Stony Brook, Stony Brook, NY 11794-3840, USA}
\begin{abstract}
In this article, we present quantum algorithms for estimating von Neumann entropy and Renyi entropy, which are crucial physical and information-theoretical properties of a given quantum state $\rho$. Although there have been existing works that achieved the same goal,  some prior developments assume the unitary that prepares the purification to the target state $\rho$. Here, we consider an alternative setting where only copies of $\rho$ are given and construct a quantum algorithm that estimates the desired entropy. Our framework can complete the given task by measuring a small number of ancilla qubits without directly measuring the system, and that it achieves significant improvement over prior relevant developments. For example, for the Renyi entropy of the order of non-integral $\alpha$, our method achieves almost power-of-two improvement in sample complexity with respect to the rank of the given state and almost a power-of-two improvement in error tolerance compared with the work by Wang et al. [Phys. Rev. Applied 19, 044041 (2023)]. 
\end{abstract}
\maketitle

\section{Introduction }
\label{sec: introduction}
Quantum computation has emerged as new computational frontier~\cite{feynman2018simulating, deutsch1985quantum, deutsch1992rapid, lloyd1996universal, shor1999polynomial, grover1996fast, berry2007efficient,berry2012black,berry2014high,berry2015hamiltonian, huang2020predicting, huang2021efficient, low2017optimal,low2019hamiltonian, arute2019quantum, preskill2018quantum, preskill2018simulating}.   Early proposals have suggested that, by utilizing the intrinsic features of quantum mechanics, information could be stored and processed in a peculiar manner, potentially having the capability to overcome computational hurdles in classical computation. At the core of quantum computation is the quantum algorithm, where a procedure is executed in a programmable manner, utilizing quantum resources that output the solution to any given problem. Beginning with the Deutsh algorithm \cite{deutsch1985quantum}, followed by the breakthroughs of Shor's factorization algorithm \cite{shor1999polynomial} and Grover's search algorithm \cite{grover1996fast}, countless quantum algorithms have been invented to tackle a wide array of problems, stemming from a diverse domain of computation. Notable examples of these subsequent developments include quantum linear solver \cite{harrow2009quantum, childs2017quantum}, quantum data fitting \cite{wiebe2012quantum}, quantum walks \cite{ambainis2007quantum, childs2010relationship}, quantum simulation algorithm \cite{berry2007efficient,berry2012black,berry2014high,berry2015hamiltonian, childs2010relationship, low2017optimal,low2019hamiltonian}, etc.  

Recently, it has been realized that \cite{gilyen2019quantum} there is a common structure underlying many of them. More specifically, a unified and possibly simplified description of all quantum algorithms exists, namely, the quantum singular value transformation framework (QSVT). The fact that the quantum singular value transformation framework allows the unification of quantum algorithms has opened up a new and exciting exploration revenue for subsequent potential development. A meaningful question emerged: In addition to providing a common ground for quantum algorithms, can quantum singular value transformation allow the power of a quantum computer to be more exploitable? In other words, it is of great interest to examine whether or not QSVT allows any stronger quantum algorithm and also constitutes a quantum algorithm that could solve any open challenges thus far. In Ref.~\cite{nghiem2022quantum}  the authors proposed a quantum algorithm to find the largest eigenvalue, based on the classical power method, with quadratic speed-up compared to the classical algorithm. In subsequent constructions, \cite{nghiem2023improved,nghiem2023improved1}, they showed that QSVT could dramatically accelerate the execution of the quantum power method in \cite{nghiem2022quantum}, and hence improve from quadratic to superpolynomial speedup. Additionally, in ~\cite{nghiem2023improved}, the authors also showed that using tools from QSVT, such as block encoding operators and related arithmetic operations (linear combination, multiplication), the quantum gradient descent algorithm first proposed in \cite{rebentrost2019quantum} could also be substantially sped up. In the work \cite{gilyen2022quantum}, the authors showed that the QSVT technique could be used to implement the Petz recovery channel, which is a vital tool in quantum information science.  

In this work, inspired by the aforementioned examples, we continue the exploration of the power quantum singular value transformation. Here, we aim to leverage the techniques from quantum singular value transformation to construct a quantum algorithm that takes copies of given density matrix $\rho$, and output an estimation to its Renyi entropy and Von Neumann entropy. These quantities are non-linear functionals of $\rho$, which are more challenging to estimate compared to the linear functionals, such as the combination of some Pauli product observables. Meanwhile, measuring on copies of $\rho$ alone would not be sufficient to reveal any higher order of its function. 
In fact, both of these entropic quantities have been considered, and methods of measuring them have been worked out in various contexts~\cite{wang2023quantum,yirka2021qubit,acharya2020estimating,gilyen2019distributional,subramanian2021quantum,gur2021sublinear, goldfeld2024quantum,wang2024new}. For example, the works in \cite{gilyen2019distributional, subramanian2021quantum, gur2021sublinear, beckey2022variational,wang2024new} assume unitary access to the purification of state $\rho$, instead of direct copies of $\rho$. On the other hand, the works in~\cite{goldfeld2024quantum, beckey2022variational} rely on variational strategies, which is heuristic. The two most relevant works to us here are Refs.~\cite{wang2023quantum} and~\cite{acharya2020estimating}, where the authors also aimed to estimate the Renyi and Von Neumann entropies using copies of $\rho$. The difference between our strategy and \cite{wang2023quantum, acharya2020estimating} is the approach we use from the transformation of singular quantum values, allowing us to construct the so-called block encoding (see \ref{def: blockencode} for the definition) of $\rho$ and directly manipulate the arbitrary power of $\rho$, which is useful for estimating desired entropies. We shall show that, by adding further ancillas of small size, we can perform a unitary on the joint system, and measure the ancillas; we can use such measurement outcomes to estimate the desired quantities. More importantly, the number of copies of a given state required in our protocol is significantly less than that of \cite{wang2023quantum}, and under a particular condition, that is, when the rank of given state does not scale as the dimension of the system, the sample complexity is also much less than that of the direct measurement approach provided in \cite{acharya2020estimating}. Thus, our contribution makes a substantial improvement towards the application of the quantum algorithmic framework to physical problems of estimating entropies.

Our work is organized as follows. To improve readability, we provide an overview of our framework in the following section \ref{sec: technicalsummary}, involving an introduction to the necessary recipes, related techniques, and illustrative insights behind our method. We then proceed to the full treatment towards Renyi entropy and von Neumann entropy estimation in subsequent sections. First, in section \ref{sec: renyi}, we devote the construction for Renyi entropy of arbitrary non-negative order $\alpha$. We divide it into three parts: the first part \ref{sec: integeralpha} aims at $\alpha$ being integers; the second part \ref{sec: fractionalalpha} aims at $\alpha$ being a fraction and greater than $1$; and the last part \ref{sec: fracalphalessthan1} discussed the case where $\alpha$ being fractional and less than $1$. The reason for these divisions, as we shall see below, is mainly technical, as for different $\alpha$, the best tool required that produces the most efficient resource turns out to be different. Section \ref{sec: vonneumann} is devoted to the von Neuman entropy. We will provide two alternative ways, one based on direct manipulation of the density operator (section \ref{sec: qsvtvonneuman}), and the other based on polynomial approximation (to Von Neumann entropy, section \ref{sec: polynomialvonneumann}), to estimate the desired quantity. We conclude and discuss outlook in section \ref{sec: conclusion}, meanwhile, we have many central definitions and tools, stated as lemmas, in the appendix \ref{sec: preliminaries}.

\section{Overview of Techniques and Related Recipes}
\label{sec: technicalsummary}
This section aim\tcw{s} to provide a more technical overview as well as key recipes underlying our main construction, which will be provided in Sections \ref{sec: renyi} and \ref{sec: vonneumann}. We first define two main entropic quantities of interest:
\begin{align}
    \text{$\alpha$-order Renyi entropy  } S_{\alpha} = \frac{1}{1-\alpha} \log \Tr( \rho^\alpha) \label{eq:Renyi}, \\
    \text{von Neumann entropy } S_v = - \Tr( \rho \log \rho ). \label{eq:vonNeumann}
\end{align}
We remark that the latter, the von Neumann entropy, is actually the limit of the former $S_v=\lim_{\alpha=1} S_\alpha$. To achieve the goal of their estimation, we first need to construct a block encoding of operator $\rho$. We refer to Appendix~\ref{sec: preliminaries} for a concrete definition of block encoding, as well as key arithmetic operations involving block-encoded operators. We recall the following recipe that was first proposed in~\cite{lloyd2014quantum} and further improved in the seminal QSVT work  ~\cite{gilyen2019quantum, gilyen2022quantum}:
\begin{lemma}
\label{lemma: logU}
    Given multiple copies of $\rho$, then it is possible to construct a $\Delta$-approximated block encoding of $ \pi \rho/ 4$ using
    $$ \mathcal{O} \Big( \frac{1}{\Delta} \log \frac{1}{\Delta}   \Big)$$
    copies of $\rho$.
\end{lemma}
The factor $\pi/4$ in the above seems unconventional at first, however, as we shall see, it results from a tool in~\cite{gilyen2019quantum}, in particular, its Lemma 70, the polynomial approximation of $2\sin^{-1}(x)/\pi$ and its Corollary 71.
Next, using the density matrix exponentiation method in~\cite{lloyd2013quantum}, one can simulate the unitary operator $\exp(-i\rho t)$. Then we also have the following lemma from \cite{gilyen2019quantum}:
\begin{lemma}[Logarithmic of Unitary, Corollary 71 in \cite{gilyen2019quantum}]
\label{lemma: logarithmicofunitary}
    Suppose that $U = \exp(-iH)$, where $H$ is a Hamiltonian of norm at most $1/2$. Let $\epsilon \in (0,1/2]$, then we can implement an $\epsilon$-approximated block encoding of $\pi H /2$ (see further definition \ref{def: blockencode}) with $\mathcal{O}(\log \frac{1}{\epsilon} )$ uses of controlled-U and its inverse, using $\mathcal{O}(\log \frac{1}{\epsilon})$ two-qubit gates and using a single ancilla qubit. 
\end{lemma}
As $\rho/2$'s norm is at most 1/2, we first construct the operator $\exp(-i \rho/2)$ using the density matrix exponentiation technique proposed in \cite{lloyd2013quantum}. Then, applying the above lemma, one can carry out the transformation:
\begin{align}
    \begin{pmatrix}
        \exp(-i\rho/2 ) &  \cdot \\
        \cdot   & \cdot 
    \end{pmatrix} \longrightarrow  \begin{pmatrix}
        \frac{\pi\rho}{4} & \cdot \\
        \cdot & \cdot 
    \end{pmatrix}.
\end{align}
Then, we obtain the block encoding of $\pi\rho/4$. 

\smallskip\noindent
{\bf R\'enyi entropy}. Given the above block encoding, we use
Lemma~\ref{lemma: product} in Appendix~\ref{sec: preliminaries} to construct the block encoding of any non-negative integer power of $\pi\rho/4$, i.e., for $k\in \mathbb{Z}_+$ we can achieve the transformation:
\begin{align}
    \begin{pmatrix}
        \frac{\pi\rho}{4} & \cdot \\
        \cdot & \cdot 
    \end{pmatrix} \longrightarrow \begin{pmatrix}
         (\frac{\pi\rho}{4})^k & \cdot \\
        \cdot & \cdot 
    \end{pmatrix}.
\end{align} 
For $k \notin \mathbb{Z}_+$, for example, for $0 < k < 1$, Lemmas ~\ref{lemma: positive} and \ref{lemma: negative} in Appendix~\ref{sec: preliminaries}, which were derived in~\cite{gilyen2019quantum}, allow us to transform the block encoding of $\pi\rho/4$ into the following two different forms:
\begin{align}
    \begin{pmatrix}
        \frac{\pi\rho}{4} & \cdot \\
        \cdot & \cdot 
    \end{pmatrix} \longrightarrow \begin{pmatrix}
        (\frac{\pi\rho}{4})^{k} & \cdot \\
        \cdot & \cdot 
    \end{pmatrix}, \\
    \begin{pmatrix}
        \frac{\pi\rho}{4} & \cdot \\
        \cdot & \cdot 
    \end{pmatrix} \longrightarrow \begin{pmatrix}
        \frac{1}{2}(\frac{\rho}{\rho_{min}})^{-k} & \cdot \\
        \cdot & \cdot 
    \end{pmatrix},
\end{align}
where $\rho_{min}$ refers to the non-zero minimum eigenvalue of $\rho$. 

In order to estimate the two desired entropic quantities, we employ an ancillary system initialized in $\ket{\bf 0}$, and make use of the following simple procedure: by the definition of block encoding (\ref{def: blockencode}), suppose $U$ is a block encoding of some operator $A$, then we have that:
\begin{align*}
    U = \ket{ \bf{0}}\bra{ \bf{0}} \otimes A + (\cdots).
\end{align*}
In the density matrix formalism, if we apply $U$ to the state $\ket{\bf 0}\bra{\bf 0} \otimes \rho$ (for $\rho$ generally being any state), then we have that:
\begin{align}
    U (\ket{\bf 0}\bra{\bf 0} \otimes \rho) U^\dagger = \ket{\bf 0}\bra{\bf 0} \otimes (A \rho A^\dagger) + (\cdots).
    \label{7}
\end{align}
If we perform the measurement on the ancillary system, then the probability of measuring $\ket{\bf 0}$ is:
\begin{align}
    p(\ket{\bf 0}) = \Tr(  A \rho A^\dagger),
\end{align}
which can be estimated up to an additive error $\epsilon$ with $\mathcal{O}(1/\epsilon^2)$ measurements. If $A$ turns out to be a Hermitian operator, then the above probability is $p(\ket{\bf 0}) = \Tr(A^2 \rho)$. More concretely, if $A$ is $(\pi\rho/4)^k$ (as we have shown how to obtain this above), then we have that $p(\ket{\bf 0}) = \Tr( \pi^{2k} \rho^{2k+1} /4^{2k} )$. Thus, by directly manipulating the power $k$, e.g., choosing $2k+1=\alpha$ we are able to estimate $\Tr(  \rho^{\alpha} (\pi/4)^{\alpha-1} ) = (\frac{\pi}{4})^{\alpha-1}\Tr(\rho^{\alpha})$, which in turn allows us to estimate the Renyi entropy of order $\alpha$ by further taking the logarithm. We remark that as $\alpha \equiv 2k+1 \geq 1$, therefore, this method cannot estimate $\Tr( \rho^{\alpha}) $ for $0 < \alpha < 1$. Subsequently, we will show that by choosing another initial state instead of $\rho$ in equation~(\ref{7}), it is still possible to estimate the entropy in the desired range of $\alpha$. We summarize our final results and provide comparison with previous attempts in the following table. We remark that we do not make new contribution to the case where $\alpha$ being integer, therefore the table only include the regime of non-integer $\alpha$.
\begin{table}[h!]
    \centering
    \begin{tabular}{|c|c|c|c|}
    \hline
        Regime  &  This work & Ref. \cite{acharya2020estimating} &  Ref. \cite{wang2023quantum}   \\
        \hline
        \hline
        $0 < \alpha < 1$   &  $ \mathcal{O}\Big( \frac{(\dim \rho)^2}{\epsilon^2 |1-\alpha|^2\big(  \Tr \rho^2 \big)^{2(\alpha-1)} \rho_{\min}^2}  \log^5\frac{(\dim\rho)}{\epsilon |1-\alpha|\big(  \Tr \rho^2 \big)^{\alpha-1} \rho_{\min}} + \log\dim\rho\Big)   $  & $\mathcal{O} \Big( \frac{d^{2/\alpha}}{\epsilon^{2/\alpha}} \Big)$ & $ \mathcal{O}\Big(  \frac{  \sum_{k=1}^K | \binom{\alpha-1}{k}  | )^5   }{ |1-\alpha|^5  \epsilon^5 \rho_{min}^2 ( \Tr \rho^2 )^{5(\alpha-1)} }      \Big) $ \\
        \hline 
        \hline
        $1 < \alpha < 2$ & $ \mathcal{O}\Big( \frac{1}{\rho_{\min}^2} \frac{r_\rho^3   }{\epsilon^3} \log^5 \big(  \frac{r_\rho}{\rho_{\min} \epsilon} \big) + \log \dim(\rho) \Big)    $ & $\mathcal{O}( \frac{n^2}{\epsilon^2})$ & $ \mathcal{O}\Big(  \frac{  ( r_\rho^{5} \sum_{k=1}^K | \binom{\alpha-1}{k}  | )^5   }{ |1-\alpha|^5  \epsilon^5 \rho_{min}^2  }      \Big) $ \\
        \hline  
        \hline
        $2<\alpha$, $\lfloor \alpha \rfloor$ is odd  & $ \mathcal{O}\Big( \frac{ r_\rho^{3(\alpha-1)}\alpha^2}{\epsilon^3 |1-\alpha|^3} \log \big( \frac{\alpha (r_\rho)^{\alpha-1} }{|1-\alpha|\epsilon} \big) + \frac{ r_{\rho}^{3(\alpha-1)} } {|1-\alpha|^3 \epsilon^3 \rho_{\min}^2} \log^5\big({\frac{ (r_\rho)^{\alpha-1}}{|1-\alpha| \epsilon \rho_{\min}} }\big)  + \log \dim(\rho) \Big)   $ & $\mathcal{O}( \frac{n^2}{\epsilon^2})$  & $ \mathcal{O}\Big(  \frac{  ( r_\rho^{5(\alpha-1)} \sum_{k=1}^K | \binom{\alpha-1}{k}  | )^5   }{ |1-\alpha|^5  \epsilon^5 \rho_{min}^2  }      \Big) $ \\
        \hline 
        \hline
        $2<\alpha$, $\lfloor \alpha \rfloor$ is even & $ \mathcal{O}\Big(  \frac{\alpha^2 r_{\rho}^{3(\alpha-1)}}{ \rho_{\min}^{-3c}\epsilon^3 |1-\alpha|^3}\log \big( \frac{\alpha (r_\rho)^{\alpha-1} }{|1-\alpha|\epsilon} \big)  + \frac{r_\rho^{3(\alpha-1)}}{\epsilon^3 |1-\alpha|^3 \rho_{\min}^{2-4c}} \log^5 \frac{r_\rho^{\alpha-1}}{\epsilon |1-\alpha|\rho_{min}^{1-c} } + \log \dim(\rho) \Big) \Big)$ & $\mathcal{O}(\frac{n^2}{\epsilon^2}) $  & $ \mathcal{O}\Big(  \frac{  ( r_\rho^{5(\alpha-1)} \sum_{k=1}^K | \binom{\alpha-1}{k}  | )^5   }{ |1-\alpha|^5  \epsilon^5 \rho_{min}^2  }      \Big)$ \\
        \hline 
    \end{tabular}
    \caption{Table summarizing our sample complexity and complexity of Ref \cite{acharya2020estimating, wang2023quantum} in estimating Renyi entropy of non-integral order $\alpha$. }
    \label{tab: renyi}
\end{table}

\smallskip\noindent
{\bf von Neumann entropy}. With regard to the von Neumann entropy, the same line of techniques can be employed. We make use of the property that $-\log(x) = \log(1/x)  $, and in particular, for $0 <x < 1$, there exists a polynomial approximation of $\log(1/x)$ on the interval $[0,1]$. We shall show later that there are two possible approaches that  can be used to leverage such a property. The first approach, a QSVT-based approach, makes use of block-encoded operators directly; for instance, with the block encoding of $\pi\rho/4$, it allows us to transform the following:
\begin{align}
    \begin{pmatrix}
        \pi\rho/4 & \cdot \\
        \cdot & \cdot 
    \end{pmatrix} \longrightarrow \begin{pmatrix}
        \gamma \log( 4\rho^{-1}/\pi  ) & \cdot \\
        \cdot & \cdot 
    \end{pmatrix},
\end{align}
where $\gamma$ is some constant. Then Lemma \ref{lemma: positive} can be employed to transform:
\begin{align}
    \begin{pmatrix}
        \gamma \log( 4\rho^{-1}/\pi  ) & \cdot \\
        \cdot & \cdot 
    \end{pmatrix} \longrightarrow  \begin{pmatrix}
        \big(\gamma \log( 4\rho^{-1}/\pi  )\big)^{1/2} & \cdot \\
        \cdot & \cdot 
    \end{pmatrix}.
\end{align}
Then we can use the property derived in equation~(\ref{7}) to apply the above unitary to the state $\ket{\bf 0}\bra{\bf 0}\otimes \rho$, and subsequently estimate the probability of measuring $\ket{\bf 0}$, which turns out to be:
\begin{align}
    p\big(\ket{\bf 0} \big) = \gamma \Tr\Big( \rho \log \frac{4\rho^{-1}}{\pi}    \Big) = \gamma \Big( \frac{4}{\pi} + \Tr( \rho \log \rho^{-1} \big) \Big) = \frac{4}{\pi} \gamma + \gamma S_v.
\end{align}
Hence, the estimation of $p(\ket{\bf 0})$, which is done via measurement, allows us to estimate the von Neumann entropy. 

The second approach, which is a polynomial approximation approach, relies on the direct expansion $-\log(\rho) = \log(\rho^{-1}) = \sum_{i=0}^K a_i \rho^i$, where $K$ is in the truncated order, and a detailed error bound will be given later. Hence, the von Neumann entropy can be approximated as:
\begin{align}
    -\Tr \Big( \rho\log \rho \Big) \approx \sum_{i=0}^K a_i \Tr(\rho^{i+1}),
\end{align}
where $K$ is roughly $\mathcal{O}( \log \frac{1}{\epsilon})$ in order to approximate with an error of  $\epsilon$. Additionally, each term $\Tr(\rho^{i+1})$ is easy to estimate as a direct result of section~\ref{sec: integeralpha}, and hence the von Neumann entropy can be estimated this way. A summary of our framework's complexity is provided in the following table. 
\begin{table}[h!]
    \centering
    \begin{tabular}{|c|c|c|c|c|}
    \hline
         & QSVT-based approach & Polynomial approximation approach & Ref. \cite{acharya2020estimating} & Ref.\cite{wang2023quantum} \\
         \hline
         \hline
    Sample Complexity & $\Tilde{\mathcal{O}}( \frac{1}{\epsilon^4 \rho_{min}^2 } ) $ & $  \mathcal{O}( \log^4 (\frac{1}{\rho_{min}}  ) \frac{1}{\rho^2_{min}}  \frac{1}{\epsilon^2} \log^2 (\frac{1}{\epsilon}) ) $ & $\mathcal{O}( \frac{\dim(\rho)^2}{\epsilon^2} )$ & $\Tilde{\mathcal{O}}( \frac{1}{\epsilon^5 \rho_{min}^2}   )$\\
    \hline
    \end{tabular}
    \caption{A summary of sample complexity of our two approaches, and related works Ref.\cite{acharya2020estimating}, Ref.\cite{wang2023quantum}. We note that the notation $\Tilde{\mathcal{O}}$ indicates the hiding of logarithmic factors.  }
    \label{tab: vonneuman}
\end{table}

The above overview has provided us with a glimpse of how our algorithm proceeds. Now, we proceed to go into greater details of our proposed framework, with concrete description of the procedure, and related error bounds.

\section{Renyi Entropy}
\label{sec: renyi}
This section is devoted for the estimation of Renyi entropy $S_\alpha$ in Eq.~(\ref{eq:Renyi}).
We note that, in principle, $\alpha$ can be a fractional number, and below, we shall show that the complexity of estimating the above value can vary significantly depending on the value of $\alpha$, e.g., whether it is fractional or integer. 
\subsection{Integer $\alpha$}
\label{sec: integeralpha}
We first consider when $\alpha$ is a positive integer, which is the simplest and well-studied case, so we utilize prior known results. Specifically, a previous work~\cite{van2012measuring} has provided an efficient way to estimate $\Tr(\rho^\alpha)$ in this regime, using random and single-copy measurements. The number of samples required in their method is $\mathcal{O}(1/\delta^2)$ for accuracy $\delta$. 
To compute the Renyi entropy, we first take the logarithm $\log(\Tr(\rho^\alpha))$ and divide it by $1-\alpha$. As what we had is the estimation of $\Tr(\rho^\alpha)$ to some accuracy $\delta$, we need to propagate such error into $(1-\alpha)^{-1} \ \log( \Tr(\rho^\alpha) )$. Denote the resultant error as $\epsilon$ and  we need to bound the error $\epsilon$ deviating from taking the logarithm and dividing by $(1-\alpha)$, which fortunately has been worked out in Ref.~\cite{wang2023quantum}, and they show the following (including non-integer $\alpha$):
\begin{align}
    \epsilon \leq 
    \begin{cases}
        \frac{2\delta}{|1-\alpha| \  (\Tr\rho^2)^{\alpha-1} } & \text{if $\alpha \in (0,1) \cup (2,\infty)$}\\
        \frac{2\delta}{|1-\alpha| \ \Tr \rho^2  }  & \text{if $\alpha \in (1,2]. $}
    \end{cases}
\end{align} 
It is also known that $\Tr \rho^2 \geq 1/r_\rho$, where $r_\rho$ refers to the rank of $\rho$. Then we have the following relations:
\begin{align}
    \epsilon \leq 
    \begin{cases}
        \frac{2\delta(r_\rho)^{\alpha-1}  }{|1-\alpha|  } & \text{ if $\alpha \in (2,\infty)$}\\
        \frac{2\delta  r_\rho }{|1-\alpha| }  & \text{if $\alpha \in (1,2] $} \\
        \frac{ 2\delta } {|1-\alpha| (\Tr \rho^2) ^{\alpha-1}} & \text{if $\alpha \in (0,1)$.  }
    \end{cases}
    \label{eqn: errorinequality}
\end{align}
In this section, we are working in the integer regime. If $\alpha =2$, then we need to choose $\delta = \epsilon/(2 r_\rho)$. If $\alpha > 2$, then we need to choose $\delta = |1-\alpha| \epsilon/ (2 r_\rho^{\alpha-1} )$.  Therefore, in order to estimate $\alpha$-th order Renyi entropy (for $\alpha$ being an integer $>1$ of a given density matrix $\rho$ up to an additive accuracy $\delta$, the number of $\rho$ required in this method is 
$$   \mathcal{O}\Big( \alpha  \frac{ r_\rho^{2\alpha-2} }{\epsilon^2} \Big), $$
which suggests intuitively that it is efficient when $\rho$ has a low rank. We remark that we have dropped the factor $|1-\alpha|$ in the above estimation as it is greater than $1$ in the regime $\alpha \in (2,\infty)$.

\subsection{Non-integer $\alpha > 1$}
\label{sec: fractionalalpha}
For any $\alpha$, we can always break it as $\alpha = \gamma + c$, where $\gamma = (2k+1)$ ($k \in \mathbb{Z}_+$) is an odd number and $-1 < c < 1$ is a fraction having magnitude less than 1. For example, for $\alpha = 5.6$ we can write it as $\alpha = 5 + 0.6$, and if $\alpha = 6.4$ then we can write it as $\alpha = 7 - 0.6$. Given the $\Delta$-approximated block encoding of $\pi\rho/4$ (as a result of Lemma \ref{lemma: logU}), we can use Lemma~\ref{lemma: product} to construct the $k\Delta$-approximated block encoding of $(\pi \rho/4)^k$ using $k$ block encodings of such $\pi \rho/4$. Due to the accumulation of error, subsequently, in order to achieve the error $\epsilon$ for the targeted $(\pi \rho/4)^k$, we will need to rescale the error in each of the block encodings of $\pi\rho/4$ to $\Delta = \epsilon/k$.  \\

\noindent \textbf{Case 1: For positive $c$ ($\lfloor \alpha \rfloor$ being odd)}\\
In this case, we can first use Lemma~\ref{lemma: findingmin} to estimate (up to some accuracy, say $\theta$) the minimum eigenvalue of $\pi \rho/4$. Roughly speaking, the method of Lemma~\ref{lemma: findingmin} uses $\mathcal{O}\big( \log (n) + \log \frac{1}{\theta}\big)$  copies of block encoding of $\pi\rho/4$. Thus, it invokes a quantum circuit of depth 
$$\mathcal{O}\Big( \frac{1}{\theta}\log \frac{1}{\theta} \big (  \log \frac{1}{\theta}  + \log \dim \rho \big)  \Big).$$
The sample number of $\rho$ required is the same. As the error $\theta$ is independent of the remaining analysis, for example, of error $\epsilon$ that we desire for the entropy, we treat it to be some constant. For subsequent summary and convenience, we note the following:

\noindent
$\bullet$ The number of copies of $\rho$ required to estimate minimum eigenvalue $\rho_{min}$ of $\rho$ is $\mathcal{O}\Big( \log \dim(\rho) \Big)$ \\
The knowledge of such a lower bound allows us to use Lemma~\ref{lemma: positive} in the Appendix, for which we quote here,
\begin{lemma}[Positive Power Exponent \cite{gilyen2019quantum1},\cite{chakraborty2018power}]
    Given an $\Delta$-approximated block encoding of a positive matrix $A$ such that 
    $$ \frac{\Ibb}{\kappa} \leq A \leq \Ibb. $$
   Let $\Delta = \delta/ (\kappa \log^3(\frac{\kappa}{\delta})) $ and $c \in (0,1)$. Then we can implement a\sout{n} $\delta$-approximated block encoding of $A^c/2$ in time complexity $\mathcal{O}( \kappa T_A \log^2 (\frac{\kappa}{\delta})  )$, where $T_A$ is the complexity of block encoding of $A$. 
\end{lemma}
That is we can transform the $\Delta$-approximated block encoding of $\pi\rho/4$ into:
\begin{align}
    \begin{pmatrix}
       \frac{\pi \rho}{4} & \cdot  \\
      \cdot  & \cdot  \\
    \end{pmatrix} \longrightarrow 
    \begin{pmatrix}
       \frac{1}{2} (\frac{\pi \rho}{4})^{c/2} & \cdot \\
       \cdot  & \cdot  \\
    \end{pmatrix}. 
    \label{eqn:67}
\end{align}
We note that we use the above lemma with $A$ being $\pi\rho/4$, so the (both circuit and sample) complexity of producing block encoding of $\pi\rho/4$ is $\mathcal{O}\Big(\frac{1}{\Delta} \log\frac{1}{\Delta}  \Big)$, according to Lemma \ref{lemma: logU}. By setting:
\begin{align}
    \kappa = \frac{1}{\rho_{\min}}, T_A = \frac{1}{\Delta}\log\frac{1}{\Delta}, \ \Delta = \frac{\delta \cdot \rho_{\min}}{  \log^3(\frac{1}{\rho_{\min}\delta}) },
\end{align}
then the right-hand side of equation~(\ref{eqn:67}) is $\delta$-approximation of $ (\frac{\pi \rho}{4})^{c/2}$. The sample complexity for producing such a block-encoded operator is hence
$$ \mathcal{O}\Big( \frac{1}{\delta \rho_{\min}^2} \log^5(\frac{1}{\rho_{\min}\delta})  \Big).$$

Given a\sout{n} $\frac{\delta}{k}$-approximated block encoding of $\pi\rho/4$ (by virtue of Lemma \ref{lemma: logU}), we can use Lemma~\ref{lemma: product} to construct the $\delta$-block encoding of $k$ products of them, e.g., $(\pi \rho /4)^k$. This step makes use of $k$ copies of the block encoding of $\pi\rho/4$, thus incurring a total of $\mathcal{O}\big( k^2 \frac{1}{\delta} \log\frac{k}{\delta}\big)$ samples of $\rho$. Then, using $\delta$-approximated block encoding of $(\frac{\pi \rho}{4})^{c/2} $ and of $(\pi \rho /4)^k$, we then use Lemma~\ref{lemma: product} to construct the $2\delta$-approximated  block encoding of $(\frac{\pi\rho}{4}) ^{k+c/2} $. So we have: 

\noindent
$\bullet$ The sample complexity for obtaining $2\delta$-approximated block encoding of $(\frac{\pi\rho}{4}) ^{k+c/2} $ is $\mathcal{O}\Big( \frac{k^2}{\delta}  \log \frac{k}{\delta} + \frac{1}{\rho^2_{min} \delta} \log^5({\frac{1}{\delta \rho_{min}} }) \Big).$  \\

Let $U_\rho$ denote the approximate block encoding of $2\delta$ of the above operator. Then from the discussion in \sout{previous} Section~\ref{sec: technicalsummary}, by employing $\ket{\bf 0}\bra{\bf 0}$ as the ancillas, we have that:
\begin{align}
U_\rho (\ket{\bf 0}\bra{\bf 0}\otimes \rho ) U_\rho = \ket{\bf 0}\bra{\bf 0} \Big(\frac{\pi\rho}{4} \Big) ^{k+ c/2} \rho \Big(\frac{\pi\rho}{4} \Big) ^{k+ c/2} + (...) 
\end{align}
If we perform the measurement on the ancillas, the probability of obtaining $\ket{\bf 0}$ is:
\begin{align}
  p_{\bf 0}  &= \Tr( (\ket{\bf 0}\bra{\bf 0}\otimes\Ibb) \cdot ( U_\rho \ket{\bf 0}\bra{\bf 0}\otimes \rho ) U_\rho^\dagger  ) \\
  &= \Tr\Big(  \big(\frac{\pi\rho}{4} \big) ^{k+ c/2} \rho \big(\frac{\pi\rho}{4} \big) ^{k+ c/2}   \Big)  \\
  &= \big(\frac{\pi}{4} \big)^{\alpha-1} \Tr \rho^\alpha = \frac{4}{\pi} \Tr \big( \frac{\pi\rho}{4} \big)^{\alpha}.
  \label{eqn: p0}
\end{align}
Note that estimating the probability $p_{\bf 0}$ is our proxy to estimating ${\rm Tr}(\rho^\alpha)$ and it can be estimated to accuracy $\delta$ with $\mathcal{O}(1/\delta^2)$ measurements. We note further that there are two factors that contribute to the error: the approximation of the block encoding and the statistical error from the estimation $p_{\bf 0}$, and the error increases linearly. It means that the estimation of $p_{\bf 0}$ to accuracy $\delta$ (thus implying the estimation of $ \pi p_{\bf 0}/4$ to accuracy $\delta$) will translate to the estimation of $\big(\frac{\pi}{4} \big)^\alpha\Tr(\rho^{\alpha})$ with error $3\delta$, as the error from the approximation of block-encoded operator $(\pi\rho/4)^{k+c/2}$ is $2\delta$. Thus:\\

\noindent
$\bullet$ It takes $\mathcal{O}\Big( \frac{k^2}{\delta^3}  \log \frac{k}{\delta} + \frac{1}{\rho^2_{min} \delta^3} \log^5({\frac{1}{\delta \rho_{min}} }) \Big)$ copies of $\rho$ to estimate $\Tr \big( \frac{\pi\rho}{4} \big)^{\alpha}$ to accuracy $3\delta$. \\

Then the $\alpha$-order Renyi entropy can be estimated by taking the logarithm of both sides and dividing them by $1-\alpha$, i.e.: 
\begin{align}
  \frac{1}{1-\alpha}  \log\Big( \Tr \big( \frac{\pi\rho}{4} \big)^{\alpha}\Big) &= \frac{1}{1-\alpha} \Big(  \log\big( \frac{\pi}{4} \big)^\alpha + \log \big( \Tr \rho^\alpha \big)   \Big) \\
  &= \frac{1}{1-\alpha}  \log\big( \frac{\pi}{4} \big)^\alpha + S_{\alpha}.
\end{align}
Let $\epsilon$ be the desired error of estimating $ \frac{1}{1-\alpha} \log \Big( \Tr \big( \frac{\pi\rho}{4} \big)^{\alpha} \Big) $. As $ \Tr \big( \frac{\pi\rho}{4} \big)^{\alpha}$ is estimated to accuracy $3\delta$, the error propagates to $\frac{1}{1-\alpha} \log \Big( \big(\frac{\pi}{4} \big)^\alpha \Tr \rho^\alpha \Big) $ as mentioned in Eqn.~\ref{eqn: errorinequality} as following:
\begin{align}
    \epsilon \leq 
    \begin{cases}
        \frac{2\cdot 3\delta \cdot r_\rho^{\alpha-1}  }{|1-\alpha|  } & \text{ if $\alpha \in (2,\infty)$}\\
        \frac{2\cdot 3\delta \cdot r_\rho }{|1-\alpha| }  & \text{if $\alpha \in (1,2] $} \\
    \end{cases}
\end{align}
which means that we need to set $\delta$ in these two respective cases as:
\begin{align}
    \delta = \frac{\epsilon |1-\alpha |}{ 6r_\rho^{\alpha-1} } \text{\ if $\alpha \in (2,\infty)$ }, \\
    \delta = \frac{\epsilon |1-\alpha|}{6 r_\rho} \text{ \ if $\alpha \in (1,2] $ }.
\end{align}
As $ \frac{1}{1-\alpha} \log \Big( \Tr \big( \frac{\pi\rho}{4} \big)^{\alpha} \Big) = \frac{1}{1-\alpha}  \log\big( \frac{\pi}{4} \big)^\alpha + S_{\alpha}$, an $\epsilon$ accuracy of estimating $\frac{1}{1-\alpha} \log \Big( \Tr \big( \frac{\pi\rho}{4} \big)^{\alpha} \Big)  $ translates directly to $\epsilon$ accuracy of $S_\alpha$. So the total number of copies of $\rho$ required to estimate $\log \big( \Tr \rho^\alpha \big)$ to accuracy $\epsilon$ is: 
\begin{align}
\mathcal{O}\Big( \frac{k^2}{\delta^3}  \log \frac{k}{\delta} + \frac{1}{\rho^2_{\min} \delta^3} \log^5\big({\frac{1}{\delta \rho_{\min}} }\big)  \Big) = \mathcal{O}\Big( \frac{ r_\rho^{3(\alpha-1)}k^2}{\epsilon^3 |1-\alpha|^3} \log \big( \frac{k (r_\rho)^{\alpha-1} }{|1-\alpha|\epsilon} \big) + \frac{ r_{\rho}^{3(\alpha-1)} } {|1-\alpha|^3 \epsilon^3 \rho_{\min}^2} \log^5\big({\frac{ (r_\rho)^{\alpha-1}}{|1-\alpha| \epsilon \rho_{\min}} }\big)  \Big). 
\end{align}
We remark further that $2k+1= \lfloor \alpha \rfloor $, and thus $k = \mathcal{O}(\alpha)$. Substituting it to the above, we have the sample complexity in this case where $\lfloor \alpha \rfloor$ being an odd number and greater than $1$ is:
\begin{align}
\mathcal{O}\Big( \frac{ r_\rho^{3(\alpha-1)}\alpha^2}{\epsilon^3 |1-\alpha|^3} \log \big( \frac{\alpha (r_\rho)^{\alpha-1} }{|1-\alpha|\epsilon} \big) + \frac{ r_{\rho}^{3(\alpha-1)} } {|1-\alpha|^3 \epsilon^3 \rho_{\min}^2} \log^5\big({\frac{ (r_\rho)^{\alpha-1}}{|1-\alpha| \epsilon \rho_{\min}} }\big)  \Big).
\end{align}
For $\alpha \in (1,2]$, which means that $\lfloor \alpha \rfloor =1$ and $k=0$, we have $ \delta = \frac{\epsilon |1-\alpha|}{6 r_\rho} = \frac{\epsilon}{6 r_\rho}$, so the total complexity is:
\begin{align}
\mathcal{O}\Big( \frac{1}{\rho^2_{\min} \delta^3} \log^5\big({\frac{1}{\delta \rho_{\min}} }\big)  \Big)  = \mathcal{O}\Big( \frac{1}{\rho_{\min}^2} \frac{r_\rho^3   }{\epsilon^3} \log^5 \big(  \frac{r_\rho}{\rho_{\min} \epsilon} \big) \Big).
\end{align}

\noindent \textbf{Case 2: For negative $c$ ($\lfloor \alpha \rfloor $ being even)}\\
For a negative value of $c$ ($-1 < c < 0)$, we need to use a different tool, e.g., Lemma.~\ref{lemma: negative}, for which we quote as following:
\begin{lemma}[Negative Power Exponent \cite{gilyen2019quantum}, \cite{chakraborty2018power}]
   Let $-1<c<0$. Given an $\Delta$-approximated block encoding of a positive matrix $A$ such that 
    $$ \frac{\Ibb}{\kappa} \leq A \leq \Ibb. $$
    Let $\Delta = \delta / ( \kappa^{1-c} (1-c) \log^3(\frac{\kappa^{1-c}}{\delta} )  )$. Then we can implement an $\epsilon$-approximated block encoding of $A^{c}/(2\kappa^{-c})$ in complexity $\mathcal{O}( \kappa T_A  (1-c) \log^2(  \frac{\kappa^{1-c}}{\delta} ) )$.
\end{lemma}
The above tool allows us to transform block-encoded operator $\pi\rho/4$ as follows:
\begin{align}
    \begin{pmatrix}
       \frac{\pi \rho}{4} & \cdot \\
       \cdot  & \cdot  \\
    \end{pmatrix} \longrightarrow 
    \begin{pmatrix}
       \frac{1}{2}  (\frac{\rho}{\rho_{\min}})^{c/2}  & \cdot  \\
       \cdot  & \cdot  \\
    \end{pmatrix}.
    \label{eqn: 52}
\end{align}
Based on the above lemma (and also Lemma \ref{lemma: logU}), if we have a $\Delta$-approximated block encoding of $\pi\rho/4$, and if we set: 
\begin{align}
  \kappa = \frac{1}{\rho_{\min}}, T_A =  \frac{1}{\Delta}\log \frac{1}{\Delta}, \    \Delta =  \frac{\delta  \rho_{min}^{1-c}}{  (1-c) \log^3 \big( 1/ (\delta\rho_{min}^{1-c}) \big) },
\end{align}
then the right-hand side of Eqn.~(\ref{eqn: 52}) is a $\delta$-approximated block encoding of $ \frac{1}{2}  (\frac{\rho}{\rho_{min}})^{c/2}$, with sample complexity being:
$$ \mathcal{O} \Big(  \frac{1}{\delta\rho_{\min}^{2-c}}  \ \log^5 \frac{1}{\delta\rho_{min}^{1-c} }  \Big).$$
where we have omitted the factor $1-c$ as it is $\mathcal{O}(1)$.

To proceed, we first use $k$ copies $\delta/k$-approximated block encodings of $\pi \rho/4$ to construct the $\delta$-approximated block encoding of their products, e.g., $(\pi\rho/4)^k$. Then from the $\delta$-block encoding of $ \frac{1}{2}  (\frac{\rho}{\rho_{min}})^{c/2}$, we construct the $2\delta$-approximated block encoding of 
$$ ( \frac{\pi \rho}{4})^k  \  \frac{1}{2}  (\frac{\rho}{\rho_{min}})^{c/2}  = \frac{1}{2} (\frac{\pi}{4})^k  (\frac{1}{\rho_{min}})^{c/2}\rho^{k+c/2}.$$
As it takes $\mathcal{O}\Big( \frac{k}{\delta} \log\frac{k}{\delta}  \Big)$ copies of $\rho$ to construct $\delta/k$-approximated block encodings of $\pi\rho/4$, and further $k$ copies of such block encodings to construct the $\delta$-approximated block encoding of $(\pi\rho/4)^k$, we have the following statement:\\

\noindent
$\bullet$ It takes $\mathcal{O}\Big( \frac{k^2}{\delta}\log\frac{k}{\delta} +  \frac{1}{\delta\rho_{\min}^{2-c}}  \ \log^5 \frac{1}{\delta\rho_{min}^{1-c} }   \Big)$ copies of $\rho$ to construct the $2\delta$-approximated block encoding of $\frac{1}{2} (\frac{\pi}{4})^k  (\frac{1}{\rho_{min}})^{c/2}\rho^{k+c/2} $.  \\

Again, we abuse the notation $U_\rho$ to denote such unitary for block encoding. Then, we repeat the same procedure as before, in the positive-$c$ case (see equation~\ref{eqn: p0} plus the step before that), and we arrive at the probability of measuring $\ket{\bf 0}$ in the ancilla:
\begin{align}
    p_{\bf 0} = \frac{1}{4} (\frac{\pi}{4})^{2k} (\frac{1}{\rho_{min}})^c \Tr(\rho^{2k+1+c}) = \frac{1}{4} (\frac{\pi}{4})^{2k} (\frac{1}{\rho_{min}})^c \Tr(\rho^\alpha),
\end{align}
which can be estimated to accuracy $\delta$ using $\mathcal{O}(1/\delta^2)$ measurements. The error $\delta$ of $p_{\bf 0}$ accumulated linearly with the error $2\delta$ of the block-encoded operator 
$$\frac{1}{2} (\frac{\pi}{4})^k  (\frac{1}{\rho_{min}})^{c/2}\rho^{k+c/2}. $$
Thus, the quantity $\frac{1}{4} (\frac{\pi}{4})^{2k} (\frac{1}{\rho_{min}})^c \Tr(\rho^\alpha)$ can be estimated with accuracy $3\delta$ using $\mathcal{O}\Big( \frac{k^2}{\delta^3}\log\frac{k}{\delta} +  \frac{1}{\delta^3\rho_{\min}^{2-c}}  \ \log^5 \frac{1}{\delta \rho_{min}^{1-c} }   \Big) $ copies of $\rho$. 

We note that $\frac{1}{4} (\frac{\pi}{4})^{2k} \frac{1}{\rho_{min}^c} \Tr(\rho^\alpha) = \frac{1}{\rho_{min}^c} \frac{1}{\pi} \Tr\Big(  \big( \frac{\pi\rho}{4}\big)^\alpha  \Big)$, where we used $2k+1 = \alpha$. As we can estimate it with an accuracy $3\delta$, by scaling $\delta \longrightarrow \delta \frac{1}{4\rho_{min}^c}$, we can estimate the quantity $\Tr\Big(  \big( \frac{\pi\rho}{4}\big)^\alpha  \Big) $ to accuracy $3\delta$. So we have that:\\

$\bullet$ $\mathcal{O}\Big( \frac{k^2\rho_{\min}^{3c}}{\delta^3}\log\frac{k}{\delta} +  \frac{\rho_{\min}^{3c}}{\delta^3\rho_{\min}^{2-c}}  \ \log^5 \frac{1}{\delta\rho_{min}^{1-c} }   \Big ) = \mathcal{O}\Big( \frac{k^2}{\rho_{\min}^{-3c}\delta^3}\log\frac{k}{\delta} +  \frac{1}{\delta^3\rho_{\min}^{2-4c}}  \ \log^5 \frac{1}{\delta\rho_{min}^{1-c} }   \Big ) $ copies of $\rho$ suffices for estimating $\Tr\Big(  \big( \frac{\pi\rho}{4}\big)^\alpha  \Big) $ with accuracy $3\delta$.  \\

From $\Tr\Big(  \big( \frac{\pi\rho}{4}\big)^\alpha  \Big)$, taking the log and dividing both sides by $1-\alpha$, we have that:
\begin{align}
    \frac{1}{1-\alpha} \log\Big(\frac{1}{4} (\frac{\pi}{4})^{2k} (\frac{1}{\rho_{min}})^c \Tr(\rho^\alpha) \Big) &= \frac{1}{1-\alpha} \log \Big(  \frac{1}{4} (\frac{\pi}{4})^{2k} (\frac{1}{\rho_{min}})^c  \Big) + \frac{1}{1-\alpha} \log( \Tr(\rho^\alpha) )  \\
    &= \frac{1}{1-\alpha} \log \Big(  \frac{1}{4} (\frac{\pi}{4})^{2k} (\frac{1}{\rho_{min}})^c  \Big) + S_\alpha.
\end{align}
Similar to previous positive-$c$ case, let $\epsilon$ be a desired error for estimating $ \frac{1}{1-\alpha} \log\Big(\frac{1}{4} (\frac{\pi}{4})^{2k} (\frac{1}{\rho_{min}})^c \Tr(\rho^\alpha) \Big) $, then we need to set:
\begin{align}
    \delta = \frac{\epsilon|1-\alpha|}{6 r_\rho^{\alpha-1}}.
\end{align}
We remark that the ability to estimate
$$ \frac{1}{1-\alpha} \log \Big(  \frac{1}{4} (\frac{\pi}{4})^{2k} (\frac{1}{\rho_{min}})^c  \Big) + S_\alpha $$
to accuracy $\epsilon$ translates directly to the estimation of $S_{\alpha}$ with the same accuracy $\epsilon$. Thus, we arrive at the final complexity to estimate the desired entropy with precision $\epsilon$:
$$ \mathcal{O}\Big(  \frac{\alpha^2 r_{\rho}^{3(\alpha-1)}}{\rho_{\min}^{-3c}\epsilon^3 |1-\alpha|^3}\log \big( \frac{\alpha (r_\rho)^{\alpha-1} }{|1-\alpha|\epsilon} \big)  + \frac{r_\rho^{3(\alpha-1)}}{\epsilon^3 |1-\alpha|^3 \rho_{\min}^{2-4c}} \log^5 \frac{r_\rho^{\alpha-1}}{\epsilon |1-\alpha|\rho_{min}^{1-c} }  \Big),$$
where we have used the fact that $2k = \lfloor \alpha \rfloor $, so $k =\mathcal{O}(\alpha)$. Again, the above complexity suggests that the sample complexity is most efficient if $\rho$ has a low rank. 

\subsection{Fractional $0 < \alpha < 1$}
\label{sec: fracalphalessthan1}
This is a bit trickier case as we cannot apply the same procedure to $\rho$, as it already contains power $1$ of $\rho$ and hence we cannot produce a final power less than $1$. To handle this scenario, we recall that previously, once we use Lemma \ref{lemma: logU} to construct a $\Delta$-approximated block encoding of $\pi\rho/4$, then we can use Lemma \ref{lemma: positive} to obtain a $\delta$-approximated block encoding of $\frac{1}{2}( \frac{\pi \rho}{4} )^{\alpha/2}$. Again, we abuse $U_\rho$ to denote such unitary. Then we have that:
\begin{align}
    U_\rho \ (\ket{\bf 0}\bra{\bf 0} \otimes \frac{\Ibb}{\dim (\rho)}  ) U_\rho^\dagger &= \ket{\bf 0}\bra{\bf 0}\otimes \Big(  \frac{1}{2}( \frac{\pi \rho}{4} )^{\alpha/2}  \cdot \frac{\Ibb}{\dim (\rho)} \cdot  \frac{1}{2}( \frac{\pi \rho}{4} )^{\alpha/2}   \Big)  + (...) 
\end{align}
If we measure the ancilla, the probability of measuring $\ket{\bf 0}$ is: 
\begin{align}
    p_{\bf 0}  = \Tr\Big( \frac{\pi^\alpha}{4^{\alpha+1} \dim(\rho)} \rho^\alpha  \Big) = \frac{\pi^\alpha}{4^{\alpha+1} \dim(\rho)}  \Tr(  \rho^\alpha),
    \label{eqn: 81}
\end{align} 
which can be estimated up to accuracy $\delta$ using $\mathcal{O}(1/\delta^2)$ measurements. 

However, we point out that in this case there is another simple way to reduce measurement repetition quadratically faster. The idea is that we can directly raise the power from the $\Delta$-approximated block encoding of $\pi\rho/4$ to $\delta$-approximated block encoding of $(1/2) (\pi\rho/4)^\alpha$. According to prior discussion, the sample complexity of this step is $\mathcal{O}\Big( \frac{1}{\delta\rho_{\min}^2} \log^5 \frac{1}{\delta \rho_{\min}}  \Big)$. The maximally entangled state $\Ibb/\dim(\rho)$ can be prepared by, first use $\log(\dim(\rho))$ Hadamard gates to act on $\ket{0}^{\otimes \dim \rho}$ to prepare:
\begin{align}
    H^{\otimes \dim(\rho)}  \ket{0}^{\otimes \dim (\rho)} = \frac{1}{\dim(\rho)} \sum_{i=0}^{\dim(\rho)-1} \ket{i}.
\end{align}
We then append another set of ancillas initialized in $\ket{0}^{\otimes \dim \rho}$, and use CNOT gates to transform:
\begin{align}
    \frac{1}{\dim(\rho)} \sum_{i=0}^{\dim(\rho)-1} \ket{i} \ket{0}^{\otimes \dim \rho} \longrightarrow \frac{1}{\dim(\rho)} \sum_{i=0}^{\dim(\rho)-1} \ket{i}\ket{i}.
\end{align}
By tracing out the ancilla register, we obtain $\Ibb/\dim(\rho)$. In this way, we actually obtained a purification unitary of the maximally entangled state, which is basically Hadamard gates plus $\log( \dim (\rho))$ CNOT gates. Given such ability, one can use the result of \cite{rall2020quantum} (see Lemma 5 in \cite{rall2020quantum}), allowing us to estimate the quantity, 
$$ \Tr \Big(  \frac{1}{2} \big( \frac{\pi\rho}{4} \big)^\alpha \frac{\Ibb}{\dim(\rho)}  \Big), $$
up to additive error $\delta$ with $\mathcal{O}(1/\delta)$ usage of block encoding of $(1/2) (\pi\rho/4)^\alpha$ (and purification unitary of $\Ibb/\dim(\rho))$. We remind that there is an approximation error $\delta$ of the block-encoded operator $(1/2) (\pi\rho/4)^\alpha$, thus the quantity that we estimate is $\delta + \delta = 2\delta$ close to the true value of $ \Tr \Big(  \frac{1}{2} \big( \frac{\pi\rho}{4} \big)^\alpha \frac{\Ibb}{\dim(\rho)}  \Big)$. The sample complexity of producing $\delta$-approximated block encoding of $(1/2) (\pi\rho/4)^\alpha $ in given above, so we have:\\
$\bullet$ $\mathcal{O}\Big( \frac{1}{\delta^2 \rho_{\min}^2}  \log^5\frac{1}{\delta \rho_{\min}}\Big)$ copies of $\rho$ is required to estimate $ \Tr \Big(  \frac{1}{2} \big( \frac{\pi\rho}{4} \big)^\alpha \frac{\Ibb}{\dim(\rho)}  \Big)$ up to accuracy $2\delta$. In order to have $2\delta$ approximation of $\Tr \Big( \big( \frac{\pi\rho}{4} \big)^\alpha \Big) $, we need to rescale $\delta \longrightarrow \delta/ 2\dim(\rho)$. Thus it takes $\mathcal{O}\Big( \frac{(\dim \rho)^2}{\delta^2 \rho_{\min}^2}  \log^5\frac{(\dim\rho)}{\delta \rho_{\min}}\Big)$ copies of $\rho$ in order to estimate $\Tr \Big(   \big( \frac{\pi\rho}{4} \big)^\alpha \Big)  $ to accuracy $2\delta$. \\

Taking the logarithm of $\Tr \Big(  \big( \frac{\pi\rho}{4} \big)^\alpha \Big) $ and dividing by $1-\alpha$ yields:
\begin{align}
  \frac{1}{1-\alpha}  \log\left( \Tr \Big(  \big( \frac{\pi\rho}{4} \big)^\alpha \Big) \right) &= \frac{1}{1-\alpha}\log \Big( \frac{\pi}{4} \Big)^\alpha + \frac{1}{1-\alpha} \log\Big( \Tr \rho^\alpha \Big)  \\
  &= \frac{1}{1-\alpha}\log \Big( \frac{\pi}{4} \Big)^\alpha + S_\alpha.
\end{align}
Therefore, according to Eqn. \ref{eqn: errorinequality} (the third case), we need to set:
\begin{align}
    \epsilon = \frac{2 \cdot 2\delta}{ |1-\alpha| \big(  \Tr \rho^2 \big)^{\alpha-1}} \\
    \longrightarrow \delta = \frac{ \epsilon |1-\alpha|\big(  \Tr \rho^2 \big)^{\alpha-1}}  {4}.
\end{align}
The sample complexity is then:
\begin{align}
    \mathcal{O}\Big( \frac{(\dim \rho)^2}{\epsilon^2 |1-\alpha|^2\big(  \Tr \rho^2 \big)^{2(\alpha-1)} \rho_{\min}^2}  \log^5\frac{(\dim\rho)}{\epsilon |1-\alpha|\big(  \Tr \rho^2 \big)^{\alpha-1} \rho_{\min}}\Big).
\end{align}

\subsection{Comparison to previous works}
\label{sec: comparisonRenyi}
We remark that the aforementioned cases, positive $c$ and negative $c$, didn't include the sample required to estimate the minimum eigenvalue of $\rho$, which is $\mathcal{O}\big( \log \dim \rho \big)$ as a result of Lemma \ref{lemma: findingmin}, as we pointed out previously. We include it in the following summary and comparison:\\

\smallskip\noindent (1). In case $0 < \alpha < 1$, our sample complexity is
\begin{align}
    \mathcal{O}\Big( \frac{(\dim \rho)^2}{\epsilon^2 |1-\alpha|^2\big(  \Tr \rho^2 \big)^{2(\alpha-1)} \rho_{\min}^2}  \log^5\frac{(\dim\rho)}{\epsilon |1-\alpha|\big(  \Tr \rho^2 \big)^{\alpha-1} \rho_{\min}}  + \log \dim\rho\Big)
\end{align}
The sample complexity in \cite{acharya2020estimating} is $\mathcal{O}\Big( (\dim \rho )^{2/\alpha}/\epsilon^{2/\alpha}\Big)$, and in \cite{wang2023quantum} is 
$$ 
\mathcal{O}\Big(  \frac{  \sum_{k=1}^K | \binom{\alpha-1}{k}  |^5   }{ |1-\alpha|^5  \epsilon^5 \rho_{min}^2 ( \Tr \rho^2  )^{5(\alpha-1)} }      \Big).$$
Compared to \cite{acharya2020estimating}, our method reduces the dependence on $\dim \rho$ from $(\dim \rho)^{2/\alpha} \longrightarrow (\dim \rho)^2 $, which is a power improvement. Compared to~\cite{wang2023quantum}, where we see an almost power-of-3 improvement in $1/\epsilon$. 

\smallskip\noindent  (2). In case $\alpha$ being non-integer $1 < \alpha < 2$, the number of copies of $\rho$ required in our work is 
\begin{align}
\mathcal{O}\Big( \frac{1}{\rho_{\min}^2} \frac{r_\rho^3   }{\epsilon^3} \log^5 \big(  \frac{r_\rho}{\rho_{\min} \epsilon} \big) + \log \dim(\rho) \Big) 
\end{align}
The sample complexity in \cite{acharya2020estimating} is $\mathcal{O}( n^{2}/ \epsilon^{2}  )$ and that in~\cite{wang2023quantum} is 
$$ \mathcal{O}\left(  \frac{  r_\rho^{5} \Big( \sum_{k=1}^K | \binom{\alpha-1}{k}  | \Big)^5   }{ |1-\alpha|^5  \epsilon^5 \rho_{min}^2  }      \right).$$
Our method achieves a power-of-two improvement w.r.t rank $r_\rho$ compared to~\cite{wang2023quantum}, and almost a power-of-2 improvement in the inverse of error tolerance. In the regime where $r_\rho \sim n$, our method achieves almost a quadratic speedup w.r.t $n$ compared to~\cite{acharya2020estimating}. In low-rank regimes, e.g., $r_\rho \sim \rm poly(\log(n))$, the improvement is super-polynomial.  \\

\smallskip\noindent (3). In the case $\alpha$ being non-integer and $2< \alpha$, if $c$ is positive (equivalently, $\lfloor \alpha \rfloor$ being odd), the sample complexity is 
\begin{align}
    \mathcal{O}\Big( \frac{ r_\rho^{3(\alpha-1)}\alpha^2}{\epsilon^3 |1-\alpha|^3} \log \big( \frac{\alpha (r_\rho)^{\alpha-1} }{|1-\alpha|\epsilon} \big) + \frac{ r_{\rho}^{3(\alpha-1)} } {|1-\alpha|^3 \epsilon^3 \rho_{\min}^2} \log^5\big({\frac{ (r_\rho)^{\alpha-1}}{|1-\alpha| \epsilon \rho_{\min}} }\big)  + \log \dim(\rho) \Big) 
    \end{align}
The sample complexity of \cite{acharya2020estimating} in this case is still $\mathcal{O}(n^2/\epsilon^2)$, and that of \cite{wang2023quantum} is
$$ \mathcal{O}\left(  \frac{   r_\rho^{5(\alpha-1)} \Big(\sum_{k=1}^K | \binom{\alpha-1}{k}  | \Big)^5   }{ |1-\alpha|^5  \epsilon^5 \rho_{min}^2  }      \right). $$
Compared to~\cite{wang2023quantum}, our method achieves almost a power-of-2 improvement in $1/\epsilon$ and a power-of-2 improvement in $r_\rho$. Compared to \cite{acharya2020estimating}, our method can achieve superpolynomial speedup if $r_\rho \sim \rm poly(\log(n))$, i.e., having low rank. 

\smallskip\noindent (4). In the case $\alpha$ being non-integer and $2< \alpha$, if $c$ is negative (equivalently, $\lfloor \alpha \rfloor$ being even). The sample complexity is
\begin{align}
\mathcal{O}\Big(  \frac{\alpha^2 r_{\rho}^{3(\alpha-1)}}{ \rho_{\min}^{-3c}\epsilon^3 |1-\alpha|^3}\log \big( \frac{\alpha (r_\rho)^{\alpha-1} }{|1-\alpha|\epsilon} \big)  + \frac{r_\rho^{3(\alpha-1)}}{\epsilon^3 |1-\alpha|^3 \rho_{\min}^{2-4c}} \log^5 \frac{r_\rho^{\alpha-1}}{\epsilon |1-\alpha|\rho_{min}^{1-c} } + \log \dim(\rho) \Big)
\end{align}
Similar to the previous case, for negative $c$, the complexity of \cite{acharya2020estimating} is $\mathcal{O}(n^2/\epsilon^2)$ and that of \cite{wang2023quantum} is
$$ \mathcal{O}\left(  \frac{   r_\rho^{5(\alpha-1)} \Big(\sum_{k=1}^K | \binom{\alpha-1}{k}  | \Big)^5   }{ |1-\alpha|^5  \epsilon^5 \rho_{min}^2  }      \right). $$
\sout{t}\tw{T}here is a power-of-2 improvement in $r_\rho$ compared to~\cite{wang2023quantum}, and almost power-of-2 improvement in $1/\epsilon$. Compared to~\cite{acharya2020estimating}, our method can achieve superpolynomial speedup if the rank of $\rho$ is sufficiently low, for example, $r_\rho \sim \rm poly(\log(n))$.\\

\section{Von Neumann Entropy}
\label{sec: vonneumann}
Previously, we have seen how the techniques from quantum singular value transformation could be utilized to estimate Renyi entropies. Here, we more or less extend those recipes to the von Neumann entropy.
As we can see, the difference for the von Neumann entropy case is that it contains the term $\log(\rho)$, which is not a power of $\rho$. Hence, adjustment is required in order to achieve the estimation. Below, two approaches are proposed, but both are inspired by the property that the logarithm function admits an efficient polynomial expansion. By efficiency, as we shall show explicitly later, we mean that the degree of such expansion is logarithmic to the inverse of error. 

\subsection{A Method Based on Quantum Singular Value Transformation }
\label{sec: qsvtvonneuman}
First we recall from the Lemma \ref{lemma: logU} that we are equipped with a $\Delta$-approximated block encoding of $\pi \rho/4$ (using $\frac{1}{\Delta} \log( \frac{1}{\Delta})$ copies of $\rho$). To proceed, we recall a result from \cite{gilyen2019distributional} \cite{gilyen2019quantum}:
\begin{lemma} [\cite{gilyen2019distributional} \cite{gilyen2019quantum}]
\label{lemma: logx}
 Given $\beta \in (0,1]$ and $\epsilon \in (0,1/2]$. Then on the interval $[\beta, 1]$, there exists a polynomial $P(x)$ with $-1 \leq |P(x)| \leq 1$ such that:
\begin{align}
    \Big| P(x) - \frac{\log (1/x)}{2 \log (1/\beta)}\Big| \leq \epsilon. 
\end{align}
Moreover, the degree of $P(x)$ is $\deg (P(x)) = \mathcal{O}\Big( \frac{1}{\beta} \log( \frac{1}{\epsilon} )  \Big).$
\end{lemma}
The above Lemma can be applied, to transform the $\Delta$-approximated block encoding of $\pi\rho/4 $ into the $4 \deg(P(x)) \sqrt{\Delta}$-approximated block encoding of $\frac{1}{2 \log (4/\pi\rho_{min})} \log ( \frac{4}{\pi}\rho^{-1} ) $, where again, $\rho_{min}$ is the minimum eigenvalue of $\rho$, using $2\deg (P(x))$ block encodings of $\pi \rho/4$. 

For simplicity, let $\gamma \equiv \frac{1}{2 \log (4/\pi\rho_{min})} $ and $4 \deg(P(x)) \sqrt{\Delta} \equiv \delta$. Then we can use Lemma~\ref{lemma: positive} with $c=1/2$ to transform the $\delta$-approximated block encoding of $\gamma \log \big( \frac{4\rho^{-1}}{\pi} \big)$ into the block encoding of its square root $(\gamma \log \big( \frac{4\rho^{-1}}{\pi}) \big)^{1/2}$.
According to~\ref{lemma: positive}, if 
$$
\delta = \frac{\epsilon \log \big( 4/(\pi \rho_{max}) \big)}{ 2\log (4\pi \rho_{min} ) \cdot \log^3  \frac{ 2\log (4/ \pi \rho_{min})  }{ \epsilon\log ( 4/\pi \rho_{max}) }    }, $$
\sout{T}\tw{t}hen the resulting is an $\epsilon$-approximated block encoding of $\big(\gamma \log ( \frac{4\rho^{-1}}{\pi})\big)^{1/2}$. Denote this unitary block encoding as $U$, then we have that:
\begin{align}
    U (\ket{\bf 0}\bra{\bf 0} \otimes \rho ) U^\dagger = \ket{\bf 0}\bra{\bf 0} \otimes \Big(\gamma \log  \frac{4\rho^{-1}}{\pi} \Big)^{1/2} \ \rho \ \Big(\gamma \log  \frac{4\rho^{-1}}{\pi}  \Big)^{1/2} + (...),
\end{align}
where, again, $(...)$ denotes the auxiliary part that is irrelevant after measurement. If we perform measurement on the first register, the probability of measuring $\ket{\bf 0}\bra{\bf 0}$ is:
\begin{align}
    p_{\bf 0} &= \Tr \Big(  (\ket{\bf 0}\bra{\bf 0} \otimes \Ibb) \ U (\ket{\bf 0}\bra{\bf 0} \otimes \rho ) U^\dagger  \Big) \\
    &= \Tr \Big( \big(\gamma \log  \frac{4\rho^{-1}}{\pi} \big)^{1/2} \ \rho \ \big(\gamma \log  \frac{4\rho^{-1}}{\pi} \big)^{1/2} \Big) \\
    &= \gamma \log \Big(\frac{4}{\pi} \Big) + \gamma \Tr( \rho \log \rho^{-1} ),
\end{align}
which can be estimated to accuracy $\epsilon$ via $\mathcal{O}(1/\epsilon^2)$ repetitions. Since $\gamma$ is known, then the von Neumann entropy can be estimated by simply subtracting $p_{\bf 0}$ by $\gamma \log  (4/\pi)$ and then  dividing the result by $\gamma$. 

To sum up the complexity, we recall that it takes $\mathcal{O}(\frac{1}{\Delta} \log(\frac{1}{\Delta}) $ copies of $\rho$ to construct the $\Delta$-approximated block encoding of $\pi\rho/4$. Then, we need to transform the $\Delta$-approximated block encoding of  $\pi \rho/4$ to  $4 \deg(P(x)) \sqrt{\Delta}$-approximated block encoding of  $\frac{1}{2 \log (4/\pi\rho_{min})} \log ( \frac{4}{\pi}\rho^{-1} ) $. Eventually, we transform it to the block encoding of $ \log(  \frac{4 \rho^{-1} }{\pi} )  /  2 \log( \frac{4}{\pi\rho_{min}}  ) $. If we expect to have $\epsilon$ error approximation, we need to set: 
\begin{align}
    4 \deg (P(x)) \sqrt{\Delta} =  \frac{\epsilon \log  \big(4/\pi \rho_{max}\big) }{ 2\log (4\pi \rho_{min} ) \cdot \log^3 \big( \frac{2\log (4/\pi \rho_{min})  }{ \epsilon\log ( 4/\pi \rho_{max} )} \big)  }. 
\end{align}
Using the fact that $\deg P(x) = \frac{1}{\rho_{min}} \log(1/\epsilon)$, we obtain
\begin{align}
    \Delta = \frac{ \epsilon^2 \rho_{min}^2 \log^2 4/(\pi \rho_{max})   } { 32 \log^2(  \frac{1}{\epsilon} ) \log^2 (4/ \pi \rho_{min} )  \log^6 \big( \frac{2\log (4/ \pi \rho_{min} ) }{ \epsilon\log  (4/\pi \rho_{max}) }  \big) }.
\end{align}
Eventually, we need to repeat the process $\mathcal{O}(1/\epsilon^2)$ times to estimate the desired probability, which leads to the von Neumann entropy to accuracy $\epsilon$. Therefore, the total sample complexity is: 
$$  \mathcal{O}\Big( \frac{  2\log (4/\pi \rho_{min} )}{ \log ( 4/(\pi \rho_{max})}  \cdot \log^2 \big(\frac{1}{\epsilon} \big)    \cdot \frac{32 \log^2(  \frac{1}{\epsilon} ) \log^2 (4/\pi\rho_{min} ) ) \log^6 \big( \frac{ 2\log (4/ \pi \rho_{min} )  }{ \epsilon\log ( 4/ \pi \rho_{max} ) }\big)  }{\epsilon^2 \rho_{min}^2 \log^2 ( 4/ \pi \rho_{max} )}  \frac{1}{\epsilon^2}  \Big),  $$
which is equal to
$$
\mathcal{O}\left(  \frac{ \log^3 ( 4/\pi \rho_{min}    )}{ \log^3  (  4/\pi \rho_{max} ) }   \ \frac{1}{\epsilon^4 \rho_{min}^2} \log^4(  \frac{1}{\epsilon})  \log^6  \frac{ 2\log (4/ \pi \rho_{min} )  }{ \epsilon\log ( 4/\pi \rho_{max} ) }  \right)  = \Tilde{\mathcal{O}}\Big( \frac{1}{\epsilon^4 \rho_{min}^2 } \Big),$$
where $\Tilde{\mathcal{O}}$ omits the polylogarithmical terms. In \cite{wang2023quantum}, their sample cost of estimating von Neumann entropy is  $\Tilde{\mathcal{O}} ( \frac{1}{\epsilon^5 \rho_{min}^2} )$. Therefore, our method is slightly better by a factor of $1/\epsilon$. 

\subsection{A Polynomial Approximation Approach}
\label{sec: polynomialvonneumann}
Recall that in order to estimate the von Neumann entropy, we employed Lemma~\ref{lemma: logx} to approximate $\log(x)$ as a polynomial of low degree within some interval. One can ask, what if we use Lemma~\ref{lemma: logx} to directly approximate $-\Tr(\rho \log (\rho) )=  \Tr( \rho \log( \rho^{-1} ) )$ as a polynomial and estimate those terms within such a polynomial? As the polynomial is a low degree, one can expect that we can do it efficiently. 

Now, we execute this simple idea. Recall that, from lemma \ref{lemma: logx}, we have a polynomial $P(x)$ such that:
\begin{align}
  \Big| P(x) - \frac{\log (1/x)}{2 \log (1/\beta)}\Big| \leq \epsilon ,
\end{align}
where $\beta \leq x \leq 1$ and the degree of $P(x)$ is $\deg(P(x)) = \mathcal{O}( \frac{1}{\beta} \log( \frac{1}{\epsilon}  ))$. From the above, we have that:
\begin{align}
    \Big| 2\log \big( \frac{1}{\beta} \big) P(x) - \log (\frac{1}{x}) \Big| \leq 2\log (\frac{1}{\beta})\epsilon.
\end{align}
Therefore, if we re-scale $\epsilon \longrightarrow \epsilon/ 2\log(1/\beta)$, then $2\log(1/\beta) P(x)$ is an $\epsilon$ approximation of the polynomial $\log(1/x)$. As $P(x)$ is a polynomial of known coefficients, $2 \log(1/\beta) P(x)$ has known coefficients. More concretely, we write that:
\begin{align}
   2\log (\frac{1}{\beta}) P(x) = \sum_{i=0}^K a_i x^i,
\end{align}
where $a_i \in \mathbb{C}$ for $i=0,2,.., K$  are known coefficients (see~\cite{gilyen2019distributional} and~\cite{gilyen2019quantum}) and $K \in \mathcal{O}( \frac{1}{\beta} \log \frac{1}{\epsilon}  )$. Therefore, equivalently, we have:
\begin{align}
    \log \rho^{-1} \approx \Big(2\log \frac{1}{\rho_{min}} \Big) \sum_{i=0}^K a_i \rho^i,
\end{align}
as well as
\begin{align}
    \rho \log (\rho^{-1}) \approx \Big(2\log   \frac{1}{\rho_{min}} \Big)  \sum_{i=0}^K a_i \rho^{i+1}.
\end{align}
Therefore, we arrive at
\begin{align}
    \Tr\big( \rho \log \rho^{-1}  \big) \approx  \Big(2\log \frac{1}{\rho_{min}} \Big)  \sum_{i=0}^K a_i \Tr ( \rho^{i+1}).
\end{align}
As $i$ being an integer, in section~\ref{sec: integeralpha}, we have mentioned that~\cite{van2012measuring} provides a simple method to estimate $\Tr( \rho^{i+1})$ for any integer $i$ using random measurement, with complexity $\mathcal{O}(1/\delta^2)$. As we have $K$ terms in the expansion, we need to account for the accumulation of errors, which is $\mathcal{O}( \log(\frac{1}{\rho_{min}}) K \delta)$. If we wish the error to be $\epsilon$, we need to choose $\delta = \epsilon/(K  \log 1/\rho_{min}   )$. Therefore, the total sample complexity is 
$$ \mathcal{O}\Big(  \log^2 \big( \frac{1}{\rho_{min}} \big) \frac{K^2}{\epsilon^2} \Big)  = \mathcal{O} \Big( \log^4 \big(\frac{1}{\rho_{min}}  \big) \frac{1}{\rho^2_{min}}  \frac{1}{\epsilon^2} \log^2 \frac{1}{\epsilon} \Big).$$
Recall that the sample complexity of our previous method, the method in~\cite{wang2023quantum} and that in~\cite{acharya2020estimating} are, respectively: 
$$ \Tilde{\mathcal{O}}\Big( \frac{1}{\epsilon^4 \rho_{min}^2 } \Big),\  \Tilde{\mathcal{O}} \Big( \frac{1}{\epsilon^5 \rho_{min}^2}   \Big),\,  \mathcal{O}\Big( \frac{\dim (\rho)^2}{\epsilon^2} \Big), $$ 
where $\Tilde{\mathcal{O}}$ refers to polylogarithmical terms being hidden from complexity presence. Clearly, our polynomial approximation-based approach achieves the same complexity with respect to $\rho_{min}$ and a power-of-3 improvement in reciprocal of error tolerance $\epsilon$ compared to the previous method ~\cite{wang2023quantum} (based on direct QSVT). Compared to~\cite{acharya2020estimating}, our method does not depend on the dimension of the system but only on the minimum eigenvalue of the system. Therefore, the improvement with respect to~\cite{acharya2020estimating} depends on how $\rho_{min}$ behave. In particular, if $\rho_{min}$ grows as much as the reciprocal of logarithm/polylogarithm in dimension, then we have exponential/superpolynomial reduction in sample complexity, respectively. 

\section{Conclusion}
\label{sec: conclusion}
In this work, we have made progress toward the problem of estimating quantum entropies, including Renyi entropy and von Neumann entropy. We have shown that one can use multiple copies of a given state to construct the so-called unitary block encoding of such a state. This step is achievable by combining the technique introduced in quantum principal component analysis~\cite{lloyd2013quantum} with the seminal quantum singular value transformation framework~\cite{gilyen2019quantum}. Then, using simple arithmetic techniques derived from the same context, e.g., quantum singular value transformation, we can manipulate the state to any non-negative power, including both integer and non-integer power, which yields a convenient tool to deal with the given task. Then, one is able to estimate the desired quantities by first applying a unitary to the joint system, then performing measurements on the ancillary system, which directly yield the values of Renyi entropy and von Neumann entropy. Compared with previous works in similar settings, for example,  Ref.~\cite{acharya2020estimating}, our work removes the dependence on the dimension of the system. Additionally, in comparison with~\cite{wang2023quantum}, we have achieved a major power improvement in rank as well as the error tolerance. Our improvement in this particular nonlinear regime has added a prominent example supporting the proposal in~\cite{lloyd2014quantum}, where the authors have put forward the idea that a quantum system can be used actively to uncover the structure of itself. In fact, we have specifically used copies of a quantum state to construct an operator of itself, and act on itself, followed by simple measurement to reveal the desired properties. As a whole, our framework has provided a better demonstration of the power of the quantum singular value transformation framework, showcasing its versatility in manipulating high-order functionals of state. Moreover, our examples also expand the scope of QSVT's application to a highly physical problem, e.g., estimating quantum entropies. It is highly motivated to expand and improve our method further in order to achieve better performance. For such a nonlinear regime, it is not known if the complexity of estimating the aforementioned entropies is optimal or even close to optimal. Therefore, it is of great interest to delve deeper to solve the last issue and possibly improve to achieve better complexity. We leave these questions for future exploration.

\section*{Acknowledgement}
This work was supported by the U.S. Department of Energy, Office of Science, National Quantum Information Science Research Centers, Co-design Center for Quantum Advantage (C2QA) under Contract No. DE-SC0012704. We also acknowledge support from the Center for Distributed Quantum Processing at Stony Brook University and a seed grant from the Provost's Office.

\bibliography{ref.bib}{}
\bibliographystyle{unsrt}

\clearpage
\newpage
\onecolumngrid
\appendix

\section{Preliminaries}
\label{sec: preliminaries}
Here, we summarize the main recipes of our work. We keep their statements brief but precise for simplicity, with their proofs/ constructions referred to in their original works.

\begin{definition}[Block Encoding Unitary]~\cite{low2017optimal, low2019hamiltonian, gilyen2019quantum}
\label{def: blockencode} 
Let $A$ be some Hermitian matrix of size $N \times N$ whose matrix norm $|A| < 1$. Let a unitary $U$ have the following form:
\begin{align*}
    U = \begin{pmatrix}
       A & \bullet \\
       \bullet & \bullet \\
    \end{pmatrix}.
\end{align*}
where $\bullet$ refers to other non-zero entry matrix blocks. We use this bullet notation to avoid confusion with $(\cdot)$ notation used in the main text that refers to $0$ entry. Then $U$ is said to be an exact block encoding of matrix $A$. Equivalently, we can write:
\begin{align*}
    U = \ket{ \bf{0}}\bra{ \bf{0}} \otimes A + \cdots,
\end{align*}
or alternatively, $A = \bra{\bf 0}\otimes \Ibb \ U \ \ket{\bf 0}\otimes \Ibb$
where $\ket{\bf 0}$ refers to the ancilla system required for the block encoding purpose. In the case where the $U$ has the form 
$$ U  =  \ket{ \bf{0}}\bra{ \bf{0}} \otimes \Tilde{A} + \cdots, $$
where $|| \Tilde{A} - A || \leq \epsilon$ (with $||.||$ being the matrix norm), then $U$ is said to be an $\epsilon$-approximated block encoding of $A$.
\end{definition}

The above definition has multiple natural corollaries. First, an arbitrary unitary $U$ block encodes itself. Suppose $A$ is block encoded by some matrix $U$, then $A$ can be block encoded in a larger matrix by simply adding ancillas (which have dimension $m$). Note that $\Ibb_m \otimes U$ contains $A$ in the top-left corner, which is a block encoding of $A$ again by definition. Further, it is almost trivial to block encode the identity matrix of any dimension. For instance, we consider $\sigma_z \otimes \Ibb_m$ (for any $m$), which contains $\Ibb_m$ in the top-left corner. 

From the above definition, suppose $\ket{\phi}$ is arbitrary state having the same dimension as $A$, we notice that:
\begin{align}
    \label{eqn: action}
    U \ket{\bf 0}\ket{\phi} = \ket{\bf 0} A\ket{\phi} + \ket{\rm Garbage},
\end{align}
where $\ket{\rm Garbage }$ generally admits the form $\sum_{j \neq \bf 0} \ket{j}\ket{\rm Garbage}_j$ is a redundant state that is completely orthogonal to $\ket{\bf 0} A\ket{\phi}$, i.e, $\bra{\rm Garbage} \ket{\bf 0} A\ket{\phi} = 0$. 

\begin{lemma}[\cite{gilyen2019quantum}]
\label{lemma: improveddme}
Let $\rho = \Tr_A \ket{\Phi}\bra{\Phi}$, where $\rho \in \mathbb{H}_B$, $\ket{\Phi} \in  \mathbb{H}_A \otimes \mathbb{H}_B$. Given unitary $U$ that generates $\ket{\Phi}$ from $\ket{\bf 0}_A \otimes \ket{\bf 0}_B$, then there exists an efficient procedure that constructs an exact unitary block encoding of $\rho$.
\end{lemma}

The proof of the above lemma is given in \cite{gilyen2019quantum} (see their Lemma 45). \\

\begin{lemma}[Block Encoding of Product of Two Matrices]
\label{lemma: product}
    Given the unitary block encoding $U_1$, $U_2$ of two matrices $A_1$ and $A_2$ using $a$ and $b$ ancillas respectively, an efficient procedure exists that constructs a unitary block encoding of $A_1 A_2$. The total ancilla usage is $a+b$.
\end{lemma}

The proof of the above lemma is also given in~\cite{gilyen2019quantum}.  \\

\begin{lemma}[\cite{camps2020approximate}]
\label{lemma: tensorproduct}
    Given the unitary block encoding $\{U_i\}_{i=1}^m$ of multiple operators $\{M_i\}_{i=1}^m$ (assumed to be exact encoding), then, there is a procedure that produces the unitary block encoding operator of $\bigotimes_{i=1}^m M_i$, which requires a single use of each $\{U_i\}_{i=1}^m$ and $\mathcal{O}(1)$ SWAP gates. 
\end{lemma}
The above lemma is a result in~\cite{camps2020approximate}. 
\begin{lemma}
\label{lemma: As}
    Given the oracle access to $s$-sparse matrix $A$ of dimension $n\times n$, then an $\epsilon$-approximated unitary block encoding of $A/s$ can be prepared with gate/time complexity $\mathcal{O}(\log n + \log^{2.5}(\frac{1}{\epsilon}))$.
\end{lemma}
This is also presented in~\cite{gilyen2019quantum}.  One can also find similar construction in Ref.~\cite{childs2017lecture}. 
\begin{lemma}
    Given unitary block encoding $\{U_i\}_{i=1}^m$ of operators $\{M_i\}_{i=1}^m$. Then, there is a procedure that produces a unitary block encoding operator of $\sum_{i=1}^m (\gamma_i/\gamma) M_i $ in complexity $\mathcal{O}(m)$, where $\gamma = \sum_i \gamma_i$. The number of extra ancillas involved are $\mathcal{O}(  \log(m) + T_U  )$, where $T_U$ is the number of ancillas required to block encode $M_i$  for all $i$. 
    \label{lemma: sumencoding}
\end{lemma}

\begin{lemma}[Scaling Block encoding]
\label{lemma: scale}
    Given a block encoding of some matrix $A$ (as in~\ref{def: blockencode}), then the block encoding of $A/p$, where $p > 1$, can be prepared with an extra $\mathcal{O}(1)$ cost.
\end{lemma}

\begin{lemma}[Positive Power Exponent \cite{gilyen2019quantum1},\cite{chakraborty2018power}]
\label{lemma: positive}
    Given an $\delta$-approximated block encoding of a positive matrix $A$ such that 
    $$ \frac{\Ibb}{\kappa} \leq A \leq \Ibb. $$
   Let $\delta = \epsilon/ (\kappa \log^3(\frac{\kappa}{\epsilon})) $ and $c \in (0,1)$. Then we can implement an $\epsilon$-approximated block encoding of $A^c/2$ in time complexity $\mathcal{O}( \kappa T_A \log^2 (\frac{\kappa}{\epsilon})  )$, where $T_A$ is the complexity of block encoding of $A$. 
\end{lemma}
We comment that in our work, as we want to construct $(\pi\rho/4)^{c/2}$, $T_A$ is the time complexity of producing block encoding of $\pi\rho/4$, which is also the number of samples required to obtain such block encoding. In particular, $T_A = \frac{1}{\delta}\log \frac{1}{\delta}$.

\begin{lemma}[Negative Power Exponent \cite{gilyen2019quantum}, \cite{chakraborty2018power}]
\label{lemma: negative}
    Given an $\delta$-approximated block encoding of a positive matrix $A$ such that 
    $$ \frac{\Ibb}{\kappa} \leq A \leq \Ibb. $$
    Let $\delta = \epsilon / ( \kappa^{1+c} (1+c) \log^3(\frac{\kappa^{1+c}}{\epsilon} )  )$. Then we can implement an $\epsilon$-approximated block encoding of $A^{-c}/(2\kappa^c)$ in complexity $\mathcal{O}( \kappa T_A  (1+c) \log^2(  \frac{\kappa^{1+c}}{\epsilon} ) )$.
\end{lemma}

\begin{lemma}[\cite{nghiem2023improved1}]
\label{lemma: findingmin}
Given the block encoding of some positive-semidefinite matrix $A$ whose eigenvalues are $\in (0,1)$, then its smallest eigenvalue can be estimated up to additive accuracy $\delta$ in time 
$$\mathcal{O}\Big( \frac{T_A}{\delta} \big( \log\frac{1}{\delta} + \frac{\log n}{2} \big)  \Big),$$
where $T_A$ is the complexity of producing the block encoding of $A$.
\end{lemma}

\end{document}